\documentclass[twocolumn,prl,a4paper,superscriptaddress]{revtex4}
\usepackage{comment}
\usepackage{color}
\usepackage{amsmath}
\usepackage{amssymb}
\usepackage{graphicx}
\usepackage{braket} 
\usepackage{esint}
\usepackage{adjustbox} 
\usepackage[utf8]{inputenc}
\usepackage{adjustbox}
\usepackage[unicode=true,pdfusetitle,
bookmarks=true,bookmarksnumbered=false,bookmarksopen=false,
breaklinks=true,pdfborder={0 0 1},backref=false,colorlinks=true]
{hyperref}
\hypersetup{
	linkcolor=red,urlcolor=blue,citecolor=blue,anchorcolor=blue}
\usepackage{breakurl}
\usepackage{ulem}
\makeatletter
\@ifundefined{textcolor}{}
{%
	\definecolor{BLACK}{gray}{0}
	\definecolor{WHITE}{gray}{1}
	\definecolor{RED}{rgb}{1,0,0}
	\definecolor{GREEN}{rgb}{0,1,0}
	\definecolor{BLUE}{rgb}{0,0,1}
	\definecolor{CYAN}{cmyk}{1,0,0,0}
	\definecolor{MAGENTA}{cmyk}{0,1,0,0}
	\definecolor{YELLOW}{cmyk}{0,0,1,0}
}

\usepackage{dcolumn}\usepackage{bm}

\makeatother

\begin{document}

\title{ Generation of perfectly entangled two and three qubits states by classical random interaction   } 
\author{Javed Akram}
\email{javedakram@daad-alumni.de}

\affiliation{Department of Physics, COMSATS University Islamabad, Pakistan}

\date{\today}
\begin{abstract} 
This study examines the possibility of finding perfect entanglers for a Hamiltonian which corresponds to several quantum information platforms of interest at the present time. However, in this study, we use  a superconducting circuit that stands out from other quantum-computing devices, especially because Transmon qubits can be coupled via capacitors or microwave cavities, which enable us to combine high coherence, fast gates, and high flexibility in its design parameters.  There are currently two factors limiting the performance of superconducting processors: timing mismatch and the limitation of entangling gates to two qubits.  In this work, we present a two-qubit SWAP and a three-qubit Fredkin gate, additionally, we also demonstrate a  perfect adiabatic entanglement generation between two and three programmable superconducting qubits. Furthermore, in this study, we also demonstrate the impact of random dephasing, emission, and absorption noises on the quantum gates and entanglement.   It is demonstrated by numerical simulation that CSWAP gate and $W$-state generation can be achieved perfectly in one step with high reliability under weak coupling conditions. Hence, our scheme could contribute to quantum teleportation, quantum communication, and some other areas of quantum information processing.  
\end{abstract} 

		
\maketitle
\section{Introduction}  \label{Sec-1}  
A number of advances have been made in creating coherent quantum information processing systems 
\cite{Monroe-2013,Abbas-2015,Awschalom-2013,Nawaz-2017,Wendin-2017,Slussarenko-2019,ArslanAnis-2021},  where the entanglement is an essential resource for quantum information processing \cite{Abbas-2009,Chuang,Ladd-2010,ABID-2022}. It is therefore very important to prepare maximally entangled states. A Greenberger-Horne-Zeilinger state and a $W$-state are two fundamental ways to entangle three qubits \cite{PhysRevA.62.062314,PhysRevA.65.032108, PhysRevA.78.012312}. Our paper examines the fastest way to create two-qubit and three-qubit entanglement states; for three-qubit states, we focus particularly on W states. A number of studies have examined how these states are created \cite{Akram-2008,Chen-2016,PhysRevLett.92.077901,PhysRevA.66.044302,Kang-2015}. A superconducting qubit can be used to create maximally entangled states in part due to their long coherence times and remarkable tuneability \cite{PhysRevA.79.052328,PhysRevA.74.064303,Wei-2015,Kang-2016,AYOUB20211353977}. An entangler gate with three qubits was used  to create W states \cite{Neeley-2010,DiCarlo-2010}. Nevertheless, it remains a challenge to determine a steady-state and optimal way for creating the $W$-state with the same quantum hardware as before.  
Transmon qubits are superconducting qubits that allow capacitively coupled qubits and permit precise time-dependent control. We can perform tailored and optimized unitary operations thanks to the high level of tunability. An adiabatic evolution can be used to achieve maximal entanglement \cite{PhysRevA.97.062343,Kang-2016}. By adiabatically interacting with the ground state and its first excited state, the desired state can be reached much more quickly than the timescale set by the energy gap. The adiabatic evolution is impossible, however, if the chosen trajectory of the time-dependent Hamiltonian includes a level crossing (or an exceedingly small gap) between the ground state and the first excited state.  There is a likelihood that the system will decohere before the adiabatic process is completed. When a level crossing occurs at an exact point on the symmetry spectrum, the initial and final states may belong to different symmetry sectors, so they cannot be transformed into one another using a Hamiltonian that preserves symmetry \cite{PhysRevA.97.062343,Kang-2016}. As a result, we employ another method to evolve the system efficiently, namely the generation of entanglement between two or three qubits by using classical randomness, somehow our proposed system is near to entanglement distillation. 
\begin{figure*}
\begin{center}     \includegraphics[width=16cm,height=5.0cm]{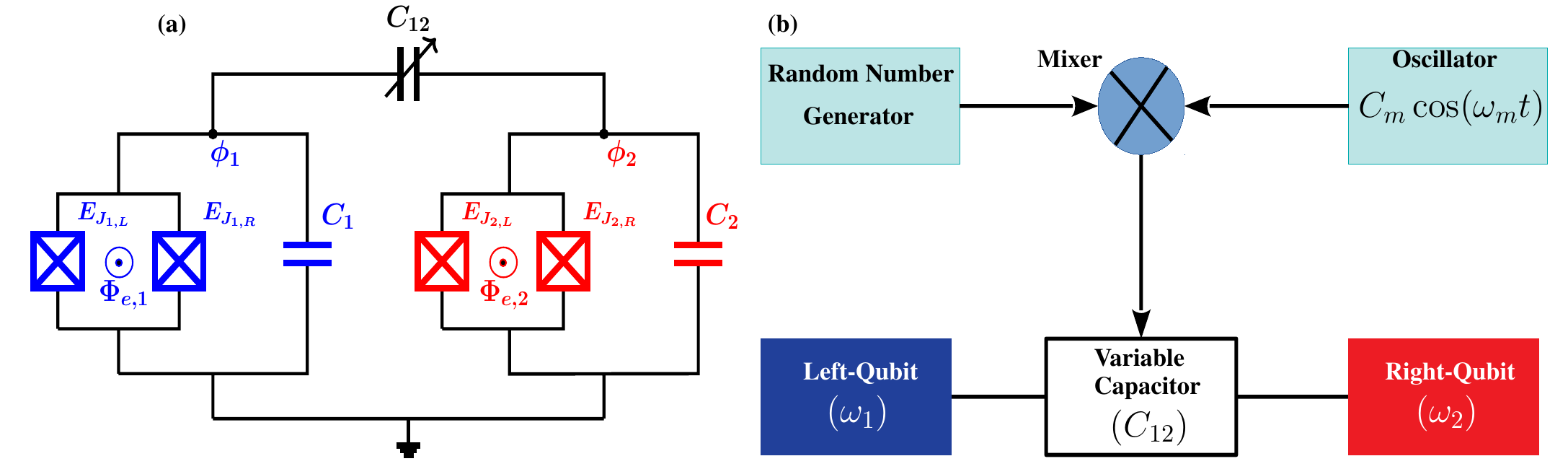}\end{center}
\caption{    (a) The circuit diagram of two connected Transmon superconducting qubits, the left, and right  Transmon qubits   have dimensionless frequencies   $\omega_{1}$ and $\omega_{2}$, respectively.  Both qubits are connected  by a variable capacitor $C_{12}$ and have a dimensionless coupling strength $g_{12}$.  (b) Block diagram of the two-qubit circuit. The variable capacitor is connected to a periodic classical oscillator and   the amplitude of the   oscillator is regulated by a random number generator.    }
\label{Fig1}
\end{figure*}

Single-qubit gates and two-qubit gates are commonly used as the basic building blocks in quantum algorithms \cite{PhysRevA.52.3457}. There are a number of elements that can be used to decompose every quantum algorithm; however, this is not a necessity.   By designing the three-qubit gates, we can reduce the circuit depth \cite{PRXQuantum.1.020304}, can improve the fidelity  of algorithms \cite{Abrams-2020}, and can tailor gates for specific problems \cite{Arute-2019,PhysRevLett.125.120504}. Therefore, several quantum algorithms require three-qubit gates, such as the Toffoli and Fredkin gates \cite{PhysRevX.8.041015,PhysRevLett.121.010501,PRXQuantum.2.040203}. As a result, composing these three-qubit gates will require considerable overhead if only standard single- and two-qubit gate sets are available \cite{PhysRevA.88.010304,Figgatt-2019}. The ability to construct native three-qubit gates at the hardware level would therefore be beneficial, but unfortunately, the types of three-body interactions that produce these gates naturally are more difficult to engineer. In this study, we also address this question and construct a two-qubit SWAP gate and a three-qubit CSWAP gate. 

In Sec. \ref{Sec-2}, we give a brief introduction to the physical coupling of qubits and build a Hamiltonian for the system.   To model the noise in our system, we use the Lindblad master equation,  we model different  types of noises such as Emission, Absorption, and Dephasing. We  examine the SWAP gate and also study the possibility of a perfect adiabatic entanglement generation between two Transmon qubits, in Sec. \ref{Sec-3}. We also investigate the effect of quantum noise on the dynamics of coupled qubits. We inspect the effect of classical random coupling strength on the two-body adiabatic entanglement.  In Sec. \ref{Sec-4}, we study the dynamics of the three coupled Transmon qubits. We inspect that the CSWAP gate is possible for this kind of scenario, we also note that  instantaneous and adiabatic  $W-$states can be generated for different schemes. In all our calculations, we have used the dimensionless parameters, however,   the dimensional values of the parameters can be taken from this Ref. \cite{li2020tunable}, where the qubit frequencies are  $\omega_1=\omega_3=4$GHz, $\omega_2=4.5$GHz, capacitors are taken as $C_1=C_2=C_3=100$fF, $C_{12}=C_{32}=1$fF, $C_{13}=0.02$fF.  Our findings and conclusions are summarized in Sec. \ref{Sec-5}.

\section{Physical Coupling of Qubits}\label{Sec-2}
To create entanglement between different quantum systems, we start with an interaction Hamiltonian that combines the degrees of freedom. Physical coupling of two transmon qubits can be achieved by a variable capacitor, as shown in Fig. \ref{Fig1}(a). The total Hamiltonian of two superconducting qubit-coupled systems takes a general form, $ H= H_0+H_{int}$, 
where the $H_0=\sum_{j=1,2}\Big[\frac{1}{2}\omega_{j}\sigma_{j}^{z}\Big]$, here to make our units dimensionless we divide the whole energy with $H= H/(\hbar \omega_c) $, where $\omega_c=E_c/\hbar$ and  $\omega_i=( \sqrt{8 E_{C_i} E_{J_i} }-E_{C_i}) $ describes qubit frequencies in dimensionless units. The dimensionless  capacitive energy relate is defined as  $E_{C_i}=e^2/(2C_i)$, and  Josephson junction energy  describes as $E_{J_i}=(\Phi_0/2\pi)^2/L_i$.  The interaction Hamiltonian $H_{int}$ couples both superconducting Transmon qubits and can be rewritten as  $ H_{int}=  g(t) \left[   \sigma^{x}_{1}\sigma_{2}^{x} +  \sigma_{1}^{z}\sigma^{x}_{2} +   \sigma^{x}_{1}\sigma^{z}_{2} + \sigma_{1}^{z}\sigma_{2}^{z} \right] $, here $\sigma^{x,z}$ defines the Pauli operators. Since both systems are Fermions, therefore there is four qubit-qubit coupling terms  \cite{kwon2021gate}. We model the dimensionless coupling constant as $g(t)=\frac{C_{12}(t)\sqrt{\omega_1\omega_2}}{2\sqrt{C_1 C_2}} $, here $C_{12}(t)$ is a time-dependent coupling capacitor.  We focus on the dynamics that are affected by the coupling term in the rotating frame, so we move to the rotating frame of reference $ H^{rot}= e^{i H_0 t  }H_{int} e^{- i H_0 t  }$,  which have both rapidly $(\omega_{1}+\omega_{2})$ and slowly $(\omega_{1}-\omega_{2})$ oscillating frequencies,

\begin{widetext}
\begin{equation}\label{Eq1} 
   H^{rot} =    g(t) \left[\sigma^{+}_{1}\sigma_{2}^{-}e^{i(\omega_{1}-\omega_{2}) t}+\sigma_{1}^{-}\sigma^{+}_{2}e^{-i(\omega_{1}-\omega_{2})t}+\sigma^{+}_{1}\sigma^{+}_{2}e^{i(\omega_{1}+\omega_{2}) t}
   +\sigma_{1}^{-}\sigma_{2}^{-}e^{-i(\omega_{1}+\omega_{2}) t}\right],  
\end{equation} 
\end{widetext}

In Eq. \ref{Eq1}, we describe $\sigma^{\pm}= (\sigma^{x} \pm \sigma^{y}) /2 $,  where $\sigma^{+}_{1}(\sigma_{2}^{+})$ is a  rising operators and $\sigma^{-}_{1}(\sigma_{2}^{-})$ is a lowering operator. We represent the dimensionless coupling constant as $g(t)=g_0 + g_m \cos(\omega_m t)$ where the amplitude of oscillator $g_m$ can have a random value between $\Re [0,1]$  with uniform distribution as shown in Fig. \ref{Fig1}(b), the capacitor dimensionless modulation frequency $\omega_m$ is another control parameter, which is known as "parametric coupling".  If $g_0=0$ and $\omega_m = |\omega_{1}-\omega_{2}|$, the fast rotation terms in Eq. \ref{Eq1} are neglected and the total dimensionless Hamiltonian of the two coupled qubits can be written as
 \begin{equation}\label{2A-3}
     H =\sum_{j=1,2}\Big[\frac{1}{2}\omega_{j}\sigma_{j}^{z}\Big] +  g_m (\sigma^{+}_{1}\sigma_{2}^{-}+\sigma_{1}^{-}\sigma^{+}_{2}).
 \end{equation}
Such an approximation is called the rotating wave approximation (RWA) \cite{shore1993jaynes}. This system of coupling qubits is a closed quantum system, which we know as an idealization of real systems. For modeling real-world scenarios, it is necessary to consider the interaction between a quantum system and the environment.  The evolution of the density matrix $ \rho $ for a closed quantum system is completely described by the dimensionless von Neumann equation $  \frac{d{\rho}}{dt}=\frac{1}{i }[H, \rho ] $ \cite{manzano2020short}, where $H$ is the total Hamiltonian of the system. However, in a real-world scenario, a qubit is an open quantum system that constantly interacts with its environment. Therefore, for precise control, this environmental phenomenon must be included in the equation of motion. To simplify the situation, we assume the Born approximation, which posits that the interaction between the qubit and the environment is so weak that the system does not affect the environment. The second is Markov's approximation, which states that the system is subject to a memory-less noise process.   The third  point is that the initial states of the system and the environment are not entangled $(\rho(t=0)=\rho_{sys}\otimes \rho_{env})$ \cite{kwon2021gate}. By considering these three assumptions, the dynamics of interacting qubits can be well described using the Lindblad-Master equation in the presence of quantum noise such as emission, absorption, and dephasing,  
\begin{equation}\label{Eq3} 
      \frac{d{\rho}}{dt}=  \frac{1}{i}[H,{\rho}]+\sum_{j=1,2} \Big( \gamma_{\downarrow} D[\sigma_{j}^{-}]\rho + \gamma_{\uparrow} D[\sigma_{j}^{+}]\rho +  \eta D[\sigma_{j}^{z}] \rho \Big)  ,
\end{equation}
    
where the super-operator defines as $D[L_j]\rho = L_j \rho L_j^\dag -\frac{1}{2} \left( L_j^\dag L_j \rho + \rho  L_j^\dag L_j  \right) $. In above equation, $\gamma_{\downarrow}$, $\gamma_{\uparrow}$, and $\eta $ are the dimensionless decay rates for Emission, Absorption, and Dephasing noise respectively \cite{cattaneo2019local}.

\begin{figure*}
\begin{center} 
\includegraphics[height=10.0cm,width=17.5cm]{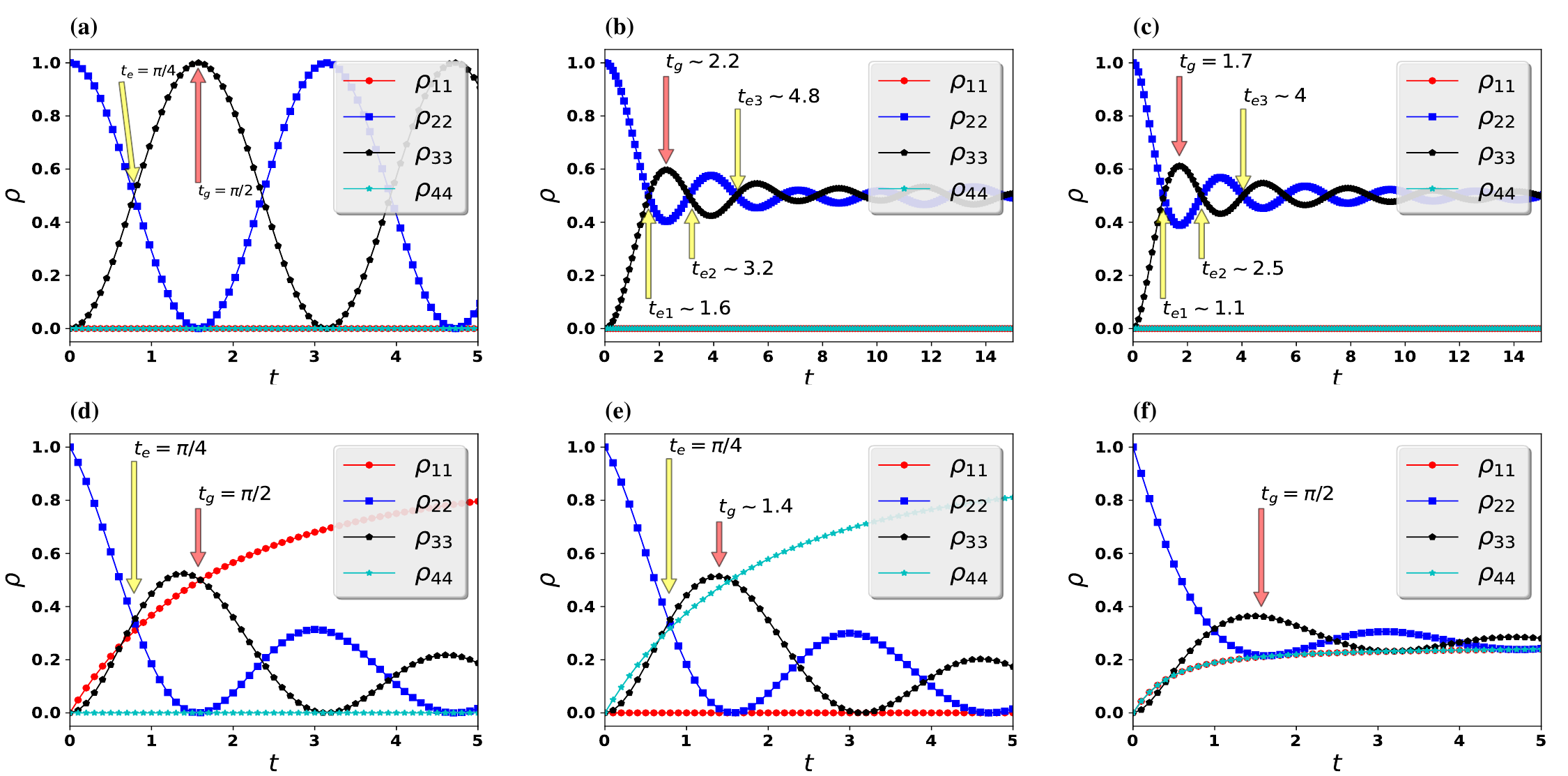} \end{center}                                       
\caption{ For different scenarios, we plot the dimensionless density ($\rho$) versus dimensionless time ($t$). The density matrix elements  can be defined as follows: red line-circle $\rho_{11} ( \ket{00} )$,  blue line-square  $\rho_{22} ( \ket{01} )$, black line-pentagon $\rho_{33} ( \ket{10} )$ and cyan line-star $\rho_{44} ( \ket{11} )$.     (a) For the ideal SWAP quantum gate, we take the dimensionless coupling strength constant $g_m=1$. The ideal SWAP gate is possible at gate time $t_g= \pi/(2 g_m) $ and two-qubit instantaneous entanglement is possible at time $t_e= \pi / (4 g_m)$.  (b) For the  random interaction strength, we use a uniformly random distribution of coupling strength $g_m=\Re [0,1]$. For the dimensionless  random dephasing   $\eta  =\Re [0,1]$(c),   emission   $\gamma_{\downarrow}  =\Re [0,1]$ (d),   absorption-noise    $\gamma_{\uparrow}  =\Re [0,1]$(e) and all noises  $\eta=\gamma_{\downarrow}  =\gamma_{\uparrow}  =\Re [0,1]$ (f).  For the last four cases (c,d,e,f), we choose the  dimensionless coupling strength   $g_m=1$ and  in order to obtain results, we perform random-disorder ensemble averages on  $ N = 1500$.}     \label{Fig2}
\end{figure*}

\section{Two-Qubit SWAP gate and perfect Entanglement }  \label{Sec-3}
Quantum gates work with qubits to uniformly manipulate the quantum states of qubits. In this subsection, we construct the SWAP gate, which is consisted of two transmon qubits, here, the SWAP quantum gate is modeled by direct variational coupling as in Fig. \ref{Fig1}.    The two qubits exchange their energy and this triggers the transition from $\ket{01}$ to $\ket{10}$. The density matrix of the system can be defined by using the basis $\rho_{11} ( \ket{00} )$, $\rho_{22} ( \ket{01} )$, $\rho_{33} ( \ket{10} )$ and $\rho_{44} ( \ket{11} )$. A perfect quantum SWAP gate $(\ket{01} \longrightarrow \ket{10}  )$ is possible at dimensionless time $t=\pi /(2 g_m )$ for an ideal case when there is no interaction with the environment as shown in Fig. \ref{Fig2}(a), in this example we have taken the initial state of the system as $\rho_{22} ( \ket{01})$. We also note that a perfect entanglement can be generated between $\ket{01} $ and $\ket{10} $  at dimensionless entanglement-time  $t=\pi /(4 g_m )$ as predicted in Fig. \ref{Fig2}(a), to plot Fig. \ref{Fig2}(a)  we have taken  $g_m=1$. We register that for the ideal case the entanglement and SWAP gate is possible at an explicit time, so to get a maximum fidelity of the SWAP gate or two-body entanglement we need to tune our system  at that certain time. If the observer is not present at this specific time then we can not achieve maximum entanglement between two states, so the question arises "can we create a steady state entanglement?".  
To achieve a steady state entanglement in this particular system, we propose a unique idea. We let the appropriate setting where the interaction strength between two Transmon qubits is randomly chosen between $\Re [0,1]$ for an ideal case, where the external environments are zero as described in Fig. \ref{Fig1}. To present our results, we perform  random-disorder ensemble averages for  $ N = 1500 $  realizations.   We determine that the fidelity of the SWAP gate decreases as the probability of the state $\rho_{33} ( \ket{10} )$ decreases to $0.6$ at dimensionless gate-time $t_g \sim 2.2$ as predicted  in Fig. \ref{Fig2}(b), this dimensionless gate-time is quite high as compared to the case Fig. \ref{Fig2}(a), where we took a  constant coupling strength, therefore, we conclude that the random interaction is not good for high fidelity SWAP gates.  However, we   note that this scenario is very useful to generate   a steady state of entanglement between state $\ket{01} $ and $\ket{10} $ for dimensionless time $t>5$ as shown in \ref{Fig2}(b). Here, we find out delay in  different entanglement generation times $t_{e1} \sim 1.6$,  $t_{e2} \sim 3.2$, and  $t_{e3} \sim 4.8$,  as described in Fig. \ref{Fig2}(b). 

Now, let's discuss another scenario, where we take interaction strength  constant $g_m=1$, but the dephasing noise  is randomly chosen between $\Re [0,1]$. We observe that the random dephasing noise can help to generate a perfectly steady state entanglement as shown in  Fig. \ref{Fig2}(c), however, the SWAP gate probability is not affected at gate-time $t_g \sim 1.7$. We also see the emission and absorption random noises effects separately on the dynamics of the two coupled Transmon qubits as shown in Fig. \ref{Fig2}(d) and  Fig. \ref{Fig2}(e), respectively. In both cases, we note the unwanted states  $\ket{00}$ and $\ket{11}$  probabilities increase for emission and absorption random noises, respectively. Therefore, we deduce  that  emission and absorption random noises are not good for the SWAP gate and also not good for steady-state entanglement generations.  Finally, for two coupled Transmon qubits, we also investigate and plot the dynamics of the system in the presence of all random noises, i.e.,  $\eta=\gamma_{\downarrow}=\gamma_{\uparrow}= \Re  [0,1]$, as shown in Fig. \ref{Fig2}(f). We determine that in this special  case, for a steady-state scenario all states become equal probably, therefore, entanglement is not possible, additionally,  we also note that the SWAP gate is also not achievable due to the combined effect of all noises. For the sake of generality, we also investigate the random interaction strength and random noises   $g_m=\eta=\gamma_{\downarrow}=\gamma_{\uparrow}= \Re  [0,1]$, at the same time, we note that for a steady-state scenario all states become equal probably as described in the previous case Fig. \ref{Fig2}(f). However, to avoid repetition, we do not present the results here.

\begin{figure*} 
\centering
\includegraphics[height=5.5cm,width=16.5cm]{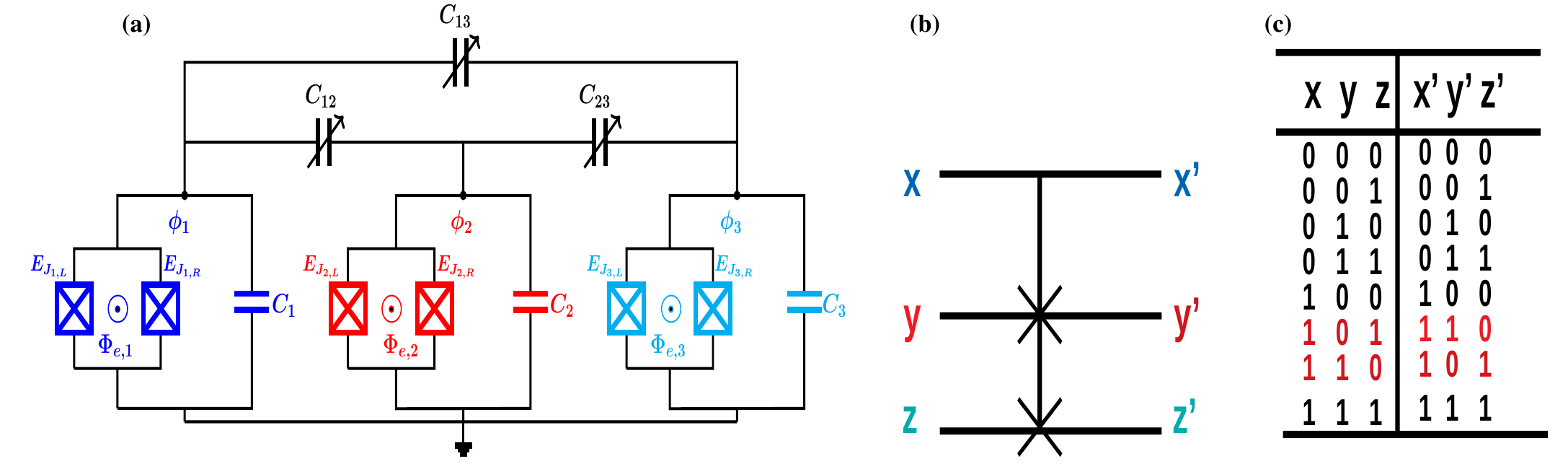}
\caption{(a) Circuit diagram of three coupled  Superconducting Transmon qubits  left ($\omega_{1}$), central ($\omega_{2}$), and right ($\omega_{3}$).  Here, $C_{12}$, $C_{23}$, and $C_{13}$ describe the variable coupling capacitors between left-central, central-right, and right-left, respectively. (b) The circuit representation of Fredkin gate or CSWAP gate. (c) The truth table of the Fredkin gate.  }
\label{Fig3}
\end{figure*}

\section{Friedkin Gate and $W-$state } \label{Sec-4}
In this section, we would like to present Friedkin gate, also known as the Controlled SWAP gate, which can be defined as  $CSWAP=\ket{0}\bra{0}  \otimes I \otimes I   + \ket{1}\bra{1} \otimes  SWAP .$      
The Friedkin Gate was first introduced by Friedkin and Toffoli in $1982$ similar to the classic three-bit logic gate \cite{fredkin1982conservative}. Friedkin Gate is not only reversible and conservative but also a Universal gate. It can be used to reproduce an AND gate and the NOT gate. The quantum version of the Friedkin gate is crucial for quantum computing, which helps in a simple implementation of Deutsch's algorithm \cite{PhysRevA.52.3489}.  Therefore, in this article, we also   implement a perfectly controlled swap gate as shown in Fig. \ref{Fig3}. We start our discussion by writing a total Hamiltonian of our three-coupled Transmon qubit system, 

\begin{widetext}
\begin{equation}\label{8-3}
     H= \sum_{j=1,2,3}\Big[\frac{1}{2}\omega_{j}\sigma_{j}^{z}\Big]+g_{12}(\sigma^{+}_{1}\sigma^{-}_{2}+\sigma_{1}^{-}\sigma_{2}^{+})+ g_{23}(\sigma^{+}_{2}\sigma^{-}_{3}+\sigma_{2}^{-}\sigma_{3}^{+})+g_{13}(\sigma^{+}_{1}\sigma^{-}_{3}+\sigma_{1}^{-}\sigma_{3}^{+}).
 \end{equation}
 \end{widetext}

Here, we have three coupled qubits with dimensionless frequencies, $\omega_{1}$(left), $\omega_{2}$(central), and $\omega_{3}$(right)  as described in Fig. \ref{Fig3}(a). All three qubits are interacting with each other, with dimensionless coupling strengths $g_{12}$(left-central), $g_{23}$(central-right), and $g_{13}$(left-right). 
Usually,  the probability of finding a system in each basis set $\{ \ket{000}, \ket{001}, \ket{010}, \ket{011}, \ket{100}, \ket{101}, \ket{110}, \ket{111} \}$ indicate by diagonal terms $\{ \rho_{00}, \rho_{11}, \rho_{22}, \rho_{33},\rho_{44},\rho_{55},\rho_{66},\rho_{77},\rho_{88}    \} $ of density matrix, respectively. 

\begin{figure*}
\begin{center} 
  \includegraphics[height=10.0cm,width=17.5cm]{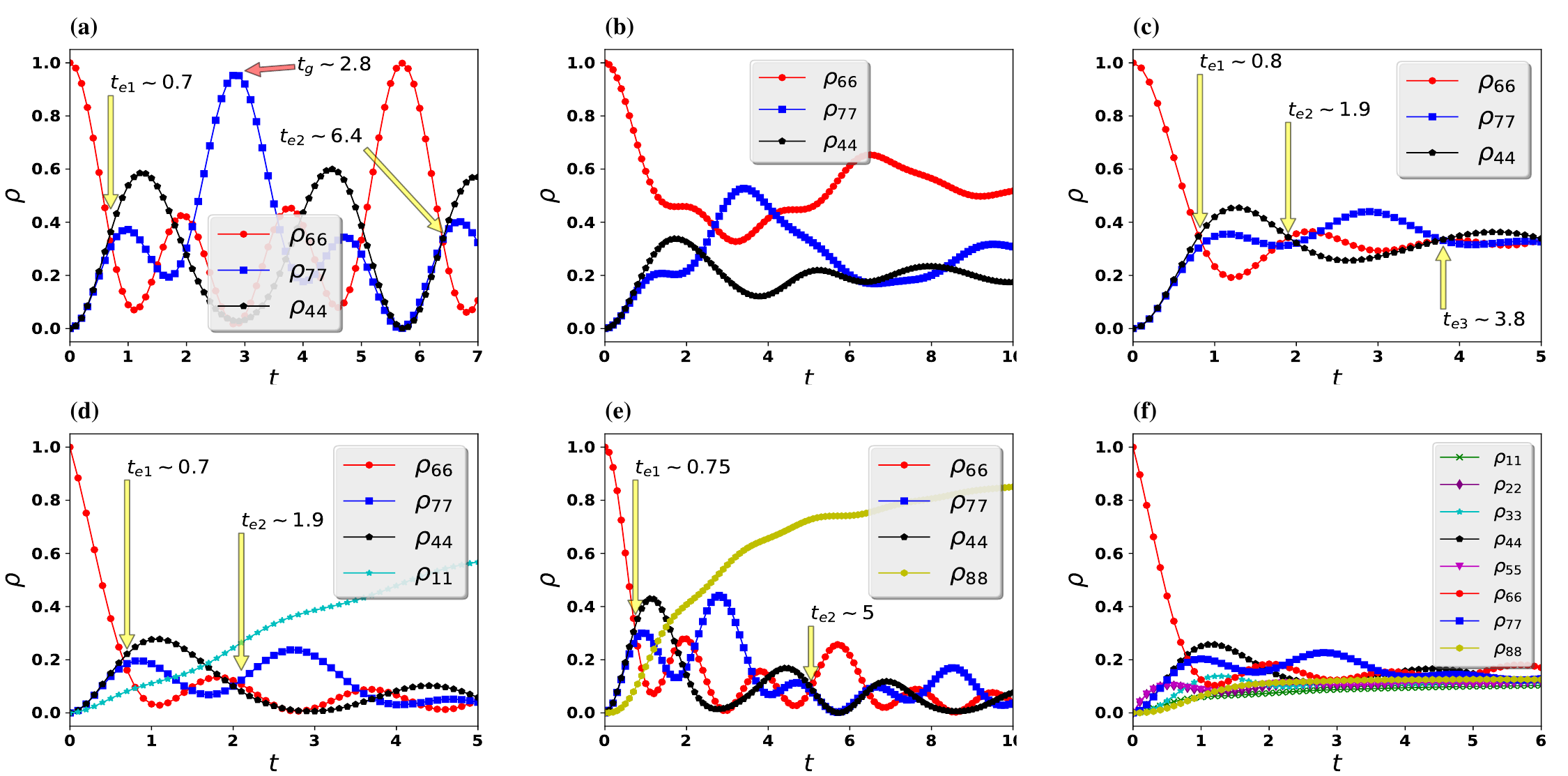} \end{center}
  \caption{We plot the dimensionless density ($\rho$) versus dimensionless time ($t$), for different scenarios.  The density matrix elements  can be defined as follows:   black line-pentagon  $\rho_{44} ( \ket{011} )$, red line-circle $\rho_{66} ( \ket{101} )$ and blue line-star $\rho_{77} ( \ket{110} )$.     (a) For the ideal CSWAP quantum gate, we take the dimensionless coupling strength constant $g_{12}=g_{23}=2 g_{13}=1$. (b) For the  random interaction strength, we use a uniformly random distribution of coupling strength $g_{12}=g_{23}=2 g_{13}=\Re [0,1]$. For the dimensionless  random dephasing   $\eta  =\Re [0,1]$(c),   emission   $\gamma_{\downarrow}  =\Re [0,1]$ (d),   absorption-noise   $\gamma_{\uparrow}  =\Re [0,1]$(e) and all noises  $\eta=\gamma_{\downarrow}  =\gamma_{\uparrow}  =\Re [0,1]$ (f).  For the last four cases (c,d,e,f), we choose the  dimensionless coupling strength   $g_{12}=g_{23}=2 g_{13}=1$ and  in order to obtain results, we perform random-disorder ensemble averages on  $ N = 150$.     } 
        \label{Fig4}
\end{figure*}

Initially, we take the maximum probability of the state $\rho_{66} (\ket{101}) $ and let the system evolve in an ideal case scenario as shown in Fig. \ref{Fig4}(a), where all the external noises are zero.  We note that the CSWAP gate is possible with a $0.98$ probability at dimensionless gate-time  $t_g \sim  2.8$ as predicted in Fig. \ref{Fig4}(a). We determine that the perfectly three-body entanglement state is also possible at dimensionless time $t_{e1} \sim 0.7 $ and $t_{e2} \sim 6.4 $ as described in Fig. \ref{Fig4}(a), for this special scenario, we  take the dimensionless qubit frequencies as $\omega_{1}=2\omega_{2}=\omega_{3}=1$  and the dimensionless coupling strengths are defined as  $g_{12}=g_{23}=2 g_{13}=1$. In literature, this perfectly three-body entanglement  is called $W-$State $\ket{W}= \big[\ket{011} +  \ket{101} + \ket{110} \big]  /\sqrt{3} $. 
Nevertheless, experimentally detecting this $W-$state is a challenge, because of external noise or due to  timing mismatch, which leads to a less measurable entangled state. To overcome this challenge, we build a technique, where the perfect entanglement can be measured with high probability. We model  the dimensionless coupling strengths as $g_{12}=g_{23}=2 g_{13}=g$, here we have taken a uniform random distribution of $g=\Re [0,1]$.  We note that by taking the random interaction strengths between all possible combinations a perfect $W-$state can be generated as shown in Fig. \ref{Fig4}(b).  We want to emphasize that all other possible states' probabilities are zero therefore we do not present them in the graph \ref{Fig4}(b). To plot this graph  \ref{Fig4}(b), we take  randomly uniformly distributed interaction strengths with an average of $150$ realizations.  To see the effect of random dephasing noise $\eta=\Re [0,1]$ on the dynamics of the system, we use the Lindblad master equation  (\ref{Eq3}). We note that the perfect equal probable three-body entangled state can be generated at the dimensionless time  $t_{e1} \sim  0.8 $, $t_{e2} \sim  1.9 $ and $t_{e3} \sim  3.8$ as shown in Fig. \ref{Fig4}(c), here we take the dimensionless coupling strengths as $g_{12}=g_{23}=2 g_{13}=1$. We also observe that the steady state $W-$state is also possible for dimensionless time $t> 4$ as described in Fig. \ref{Fig4}(c), however, we determine that the CSWAP gate is not possible for this scenario.   
After discussing the possibilities of the generation of  a perfectly three-body entangled state, now we would like to discuss the effect of emission and absorption of random noises on the dynamics of the system. We ascertain that in the presence of emission random noise $\gamma_{\downarrow}= \Re  [0,1]$, a three-body entangled state is not possible as shown in Fig. \ref{Fig4}(d)  due to the effect that $\rho_{00} (\ket{000})$ probability start increasing exponentially, other states probabilities also increase but they are not exponential increasing, therefore, we did not present them here. 


For the dimensionless random absorption  $\gamma_{\uparrow}= \Re  [0,1]$, we note that a  partially entangled state is possible at the dimensionless time $t_{e1} \sim  0.75 $ as predicted in Fig. \ref{Fig4}(e).  However, as the time increases the probability of $\rho_{88} (\ket{111})$ increases, therefore for the steady state three-body entanglement is not possible as shown in Fig. \ref{Fig4}(e). Similarly, for all random noises  $\eta=\gamma_{\downarrow}=\gamma_{\uparrow}= \Re  [0,1]$, we note that the equal probabilities of all the possible states as predicted in Fig. \ref{Fig4}(f). We can use this special scenario, for the  generation of  equal probable states, however, the three-body entanglement is out of the question as the noise effect is quite strong.

\section{Summary and Conclusion} \label{Sec-5} 
In this study, we present a design for two-qubit and three-qubit quantum gates called SWAP and CSWAP, respectively. Additionally, we also investigate the generation of adiabatic entanglement in two different systems namely, two-qubit and three-qubit states. Within the superconducting Transmon qubit architecture, we found an optimal method that entangles maximally two- and three-qubit systems in a given time. This  method  turned out to be simply based on a  classical random interaction of two- and three-qubit states. The ability to tune the time scale for these protocols leaves room for many other processes to complete before the system effect due to the decoherence, making them very useful for information processing in quantum computers. In the first part of this study, we outline the two-qubit SWAP gate, which has a maximum probability of one in the absence of the noise, however, the probability of the SWAP gate decreases to $0.6$ in the presence of the random dephasing noise. We find out that the adiabatic maximally two-qubit entangled state is possible if we model the real random interaction strength between two-qubit states over a large ensemble average. And the adiabatic perfect entanglement is also possible for the random dephasing noise, though, the entanglement is shattered in the presence of emission and absorption random noises.  We studied a detailed  method to generate a CSWAP gate which is also known as a Fredkin quantum gate, to configure this gate two coupling schemes are used, the nearest-neighbor variable coupling and the next-neighbor variable coupling between qubits. We demonstrate that for the ideal case scenario, a perfect CSWAP gate is possible with a maximum probability of $0.98$. We also noted that a  three-qubit maximally entangled $W-$state is also possible at specific times ($t_{e1},t_{e2}$) with high fidelity $ \mathcal{F} = \bra{W} \rho \ket{W}=0.33 $. Although, we achieve an adiabatic maximum entangled $W-$state, when we modeled the random interaction strength with a very high fidelity $ \mathcal{F} = \bra{W} \rho \ket{W}=0.33 $. We have also found  an adiabatic maximum entangled $W-$state in the presence of random dephasing  noise, but, the three-qubit entanglement disappeared for emission and absorption random noises. Therefore, we can safely say that the emission and absorption noises are not good for adiabatic entanglement generation. This special kind of entanglement generation technique is very close to the technique of entanglement distillation technique. All results of our research can assist experimentalists in designing quantum gates of high fidelity as well as help them to construct future quantum circuits. This concept can   be used to generate adiabatic entanglement for two-qubit and three-qubit states. Our findings are not limited to Transmon-superconducting circuits but they can also be used for Ion and Photonics circuitry.  In addition to being quite realistic, the model we propose can be experimentally implemented and all parameters can be adjusted. In the future, we would like to extend this work for time-dependent random interactions.

\section*{Acknowledgment}
J.A. gratefully acknowledges support from the National Research Program for Universities (NRPU) project No $14544$ of the Higher Education Commission of Pakistan.

\section*{Conflict of Interest}
The author declares no conflict of interest.

\section*{Data Availability Statement }
This is analytical work and no additional data is required for the production of the results. 

\section*{Keywords}
Superconducting devices, Quantum Gates,  Noisy Environment, Transmon Qubit, Coupled Lindblad master equations, Two-qubit SWAP and a three-qubit Fredkin gate, W-state generation

\bibliographystyle{apsrev4-1}

\end{document}